# Atomic imaging of 2D transition metal dihalides

Wendong Wang[1,2*], Gareth R.M Tainton[2,3*], Nick Clark[2,3†], James G. McHugh[1,2], Xue Li[1,2], Sam Sullivan-Allsop[2,3], David G. Hopkinson[4], Oldrich Cicvarek[5], Francisco Selles[1,2], Rui Zhang[2,3], Joshua D. Swindell[3], Alex Summerfield[2], David J. Lewis[3], Vladimir I Fal'ko[1,2], Zdenek Sofer[5], Sarah J. Haigh[2,3†], Roman Gorbachev[1,2†]

[1]Department of Physics and Astronomy, University of Manchester, Oxford Road, Manchester M13 9PL, United Kingdom

[2] National Graphene Institute, University of Manchester, Oxford Road, Manchester M13 9PL, United Kingdom

[3] Department of Materials, University of Manchester, Oxford Road, Manchester M13 9PL, United Kingdom

[4] Electron Physical Science Imaging Centre, Diamond Light Source, Harwell Science and Innovation Campus, Didcot, OX11 0DE, United Kingdom

[5] Department of Inorganic Chemistry, University of Chemistry and Technology Prague, Technická 5, 166 28, Prague 6, Czech Republic

* These authors contributed equally to this work.

† Corresponding author. e-mail:

roman@manchester.ac.uk, sarah.haigh@manchester.ac.uk, nick.clark@manchester.ac.uk

***Abstract:***

Transition metal di-iodides such as $FeI_2$, $NiI_2$ and $CoI_2$ are an emerging class of 2D magnets exhibiting rich and diverse magnetic behaviour, but their study at the monolayer limit has been severely hindered by fabrication challenges due to their air-sensitivity. Here, we introduce a polymer-free method for clean, rapid, and high-yield assembly of hermetically encapsulated suspended samples of air-sensitive monolayers. Applying it to di-iodides enables atomic resolution characterisation of thin samples - down to the monolayer limit - for the first time. Our imaging, combined with complementary first-principles calculations, reveals an unusually small energy barrier between alternate stable stacking polytypes in few-layer films, enabling extrinsic control of the stacking phase. We also observe stable isolated iodine vacancies that do not aggregate to form extended structures, and identify and verify the stability of the various edge configurations of thin samples. These results establish the unique structural characteristics of these materials in the thin limit, and more broadly demonstrate the utility of our transfer platform for creating atomically clean suspended vdW heterostructures.

*Introduction:*

The family of successfully isolated two-dimensional (2D) materials continues to rapidly grow, expanding the range of van der Waals (vdW) heterostructures that can be assembled layer-by-layer and therefore the scope of possible phenomena to be explored. Magnetic 2D materials[1] are relatively new members of the 2D family and offer a dramatic expansion to the potential properties achievable with vdW heterostructures. Among those, the most recent additions are the transition metal diiodides, $MI_2$, which feature rich and diverse magnetic properties which remain mostly unexplored. This includes $NiI_2$ - an extremely rare layered multiferroic, where a helical magnetic ground state[2–7] is combined with out-of-plane ferroelectric polarisation, which persists down to few-layer thickness[8]. In addition, this material has recently been shown to possess giant natural optical activity at terahertz frequencies[6] and spin-polarized edge states[7], making it one of the most interesting emerging 2D systems. Other exciting representatives of the di-iodide family include frustrated triangular magnets with rich spin dynamics and Kitaev physics, such as spin-one $FeI_2$[9] which hosts hybrid dipolar and multipolar quasiparticle excitations[10,11], and spin-half $CoI_2$[12,13].

However, like many novel atomically thin crystals[14–16], the transition metal diiodides have been found to be extremely sensitive to mild ambient conditions including exposure to oxygen, water, elevated temperature and light[15–22]. These can lead to the breakdown of the magnetic ordering or even complete disintegration of their crystalline structure[15,20,23,24], limiting the exploration of these systems to studies that employ encapsulation[25,26] or maintain inert or ultra-high vacuum (UHV) conditions[27–29]. In addition to air sensitivity, surface contamination remaining from mechanical exfoliation and transfer steps causes further problems. The contaminants can induce degradation, especially when exposed to an intense electron or laser beam, but also by preventing hermetic encapsulation, and cleaning steps to remove such contaminants[30,31] can also damage the sensitive materials and impart strain. The result is that despite their exciting magnetic behaviours, detailed atomic scale characterisation of thin di-iodide crystals is completely absent in the literature due to the challenges in fabricating atomically clean and uniform specimens. Such studies are essential to understand the magnetic properties of few-layer metal di-iodides, as well as many other novel environmentally sensitive materials, to both predict their properties and interpret experimental observations.

Here, we present an innovative platform for electron and scanning probe microscopy characterisation of air sensitive 2D materials, as well as other studies requiring suspended samples. To enable this, we have re-designed silicon nitride[32,33] ($SiN_x$) cantilevers to introduce suspended regions (Supporting information, SI, Fig. S1-S2) for subsequent imaging (SI Fig. S3-S5) and adapted them for common microscopy holders. This approach has a number of advantages over the state-of-the-art polymer transfer techniques: 1. Extremely clean 2D material surfaces and interfaces resulting from complete avoidance of organic materials and liquids; 2. On-grid fabrication avoids the need for additional specimen release onto commercial supports, improving sample integrity, quality and yield; 3. The approach enables contamination-free hermetically sealed encapsulation using stable 2D materials (graphene or hexagonal boron nitride) in an inert glove box (Ar or $N_2$) or UHV environment. Here we show this allows the first imaging of point and edge defects as well as stacking features in even the most environmentally sensitive 2D materials.

We demonstrate the efficacy of our technique by performing the first comprehensive Scanning Transmission Electron Microscopy (STEM) study of $FeI_2$, $NiI_2$ and $CoI_2$ (SI Fig. S7-S9) transition metal diiodides down to monolayer thicknesses. These $MI_2$ materials are extremely air-sensitive, with even relatively thick crystals degrading completely in less than 5 seconds on contact with air (SI Fig. S10). We use our novel stacking approach to achieve high-quality graphene encapsulation immediately after exfoliation, achieving a full hermetic seal without any interlayer contamination. The result is transition metal diiodide samples that are stable for weeks after exfoliation and can be transported between different microscopes and geographical sites. The approach is applicable to the majority of air-sensitive 2D crystals enabling atomic-resolution STEM imaging. In atomically thin $FeI_2$ and $NiI_2$ we observe unusual polytype plasticity facilitating easy stacking order switching, with potential for atomic sliding property engineering. Additionally, we also explore most common atomic vacancies, extended defects, and edge structures in these scientifically exciting yet experimentally challenging material systems.

*Results:*

The 2D transition metal dihalides were first mechanically exfoliated and then encapsulated with graphene layers on both sides using silicon nitride cantilevers in an argon glove box environment to prevent any exposure to air during the process[17,22]. The silicon nitride sample supports were prepared using standard semiconductor processes at wafer-scale, and patterned with an array of few μm diameter holes (for details see SI Section 1). This provides flexible, transparent, chemically inert, and thermally stable supports for the stacking process, see Fig. 1a. To facilitate good adhesion of the specimen and prevent electrostatic charging, the cantilevers have been coated with three metal layers (1 nm Ta, 5 nm Pt, 1.5 nm Au)[33], allowing efficient pickup of thin 2D crystals, including monolayer graphene. We find that the best yield of suspended monolayer crystals was achieved with 2 μm diameter holes, but for thicker structures the holes can be larger. To achieve encapsulation, the holey region of the $SiN_x$ cantilever was used to pick up a graphene sheet from an oxidised silicon substrate (Fig. 1b(i)). Next, a thin $MI_2$ crystal was picked up at 40°C such that it became attached to the graphene on the cantilever (Fig. 1b(ii) and Supplementary Video S1). Finally, the cantilever was used to pick up another graphene layer to complete the encapsulation (Fig. 1b(iii)). The graphene layers must be larger than the $MI_2$ crystal as even a small crack can lead to complete disintegration of the crystal, and the encapsulation procedure was performed in an argon glove box environment to prevent the $MI_2$ being exposed to air. Finally, the $SiN_x$ sample support with the graphene-$MI_2$-graphene heterostructure was released onto a custom TEM grid with a large central aperture (Fig. 1b(iv), SI Fig. S5). As seen in the higher magnification image (Fig. 1d), the encapsulated $MI_2$ crystal remains clean, stable, and visible in air.

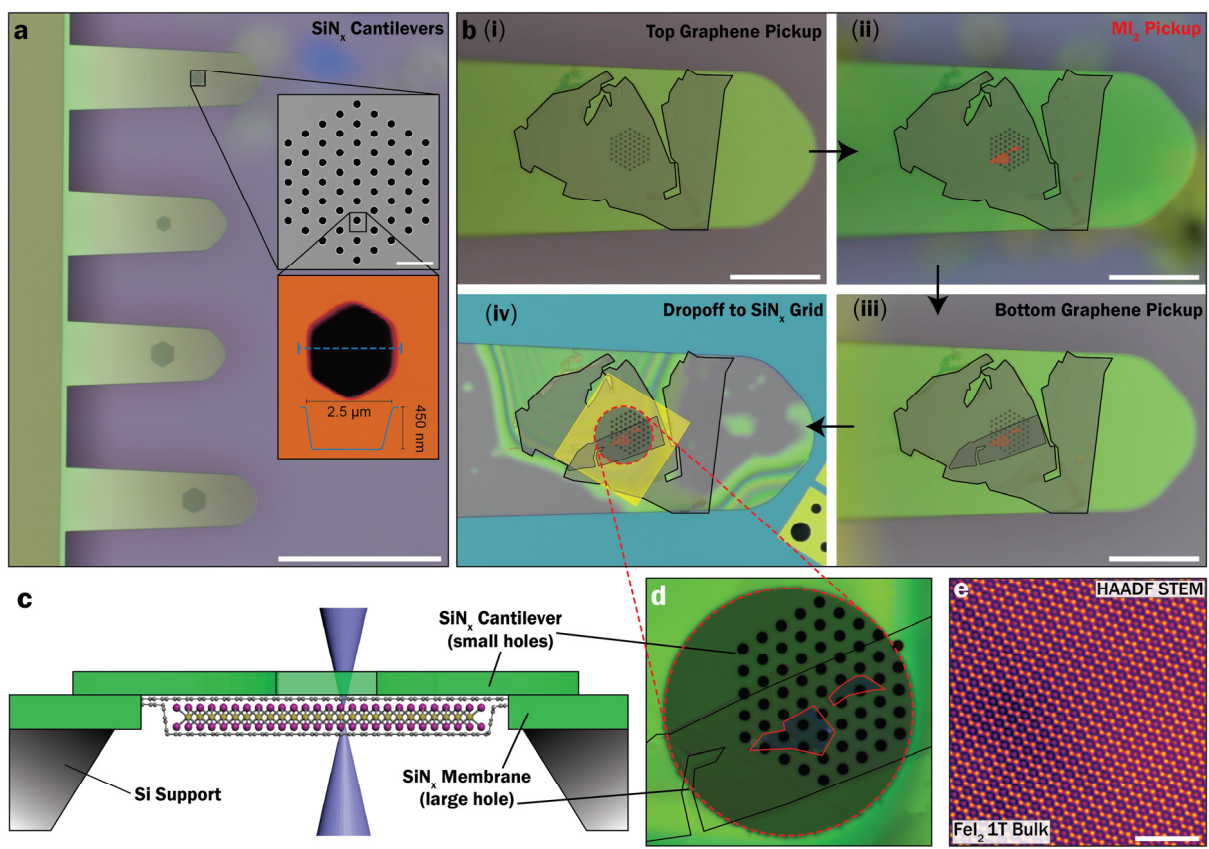

**Fig. 1 On-grid assembly of a graphene-encapsulated MI₂ heterostructures. a,** Optical image of as-fabricated holey SiN$_x$ cantilevers with different dimensions and hole arrangements. Top inset shows an SEM image of the holey region. The bottom inset shows an AFM topographic map of one hole. **b,** Steps employed to make suspended and fully encapsulated MI₂. b(i) is after the pickup of the first graphene layer, b(ii) is after the pickup of the MI₂ crystal, b(iii) is after pickup of the second graphene to complete encapsulation. b(iv) shows drop off the holey cantilever with encapsulated heterostructure onto a customised SiN$_x$ STEM support. Graphene and MI₂ crystals are outlined in black and red respectively. **c,** Schematic of the cross-section of graphene encapsulated monolayer FeI₂ on SiN$_x$ TEM grid. **d,** Magnified region from image of b(iv) showing the suspended sample. **e,** HAADF STEM image of bulk 1T-FeI₂.

### Atomic structure of MI$_2$

Individual vdW layers of MI$_2$ materials are comprised of transition metal atoms octahedrally bonded to six adjacent iodide ions (see Fig. 2g left). In bulk FeI$_2$ these layers are combined with AA-type stacking to form the 1T phase[34,35] (Fig. 2j right, SI Fig. S11), which is also the most commonly observed polytype for the majority of first row transition metal di-iodides (TiI$_2$, VI$_2$, MnI$_2$). In contrast, bulk NiI$_2$ and CoI$_2$ preferentially crystallises in the 3R phase via ABC-type stacking (Fig. 2j middle)[34,35]. These structural phases are readily distinguishable in plan view high angle annular dark field (HAADF) STEM images due to the Z-contrast variations between purely metal and purely iodine atomic columns in the 1T phase, and the presence of mixed metal and iodine atomic columns in the 3R phase. For MI$_2$ crystal thicknesses of more than 10 layers, our HAADF STEM images confirm the expected bulk stacking configurations, with bulk FeI$_2$ (Fig. 2c) matching the 1T phase, with an in plane lattice parameter of a= 4.05 ± 0.12Å, and bulk NiI$_2$ (Fig. 2f) matching the 3R phase with a= 3.92 ± 0.14Å, both in good agreement with previous observations[34,35]. Comparison of the STEM images to simulations for bilayer NiI$_2$ also reveals the expected 3R stacking in agreement with bulk material (Fig. 2e). However, images of bilayer FeI$_2$ are found to also match with 3R phase stacking (Fig. 2b, SI Fig. S12), not to the expected 1T stacking order of bulk FeI$_2$. Similar behaviour is observed in bilayer CoI$_2$ which we find in 1T phase rather than its bulk 3R form. Few-layer NiI$_2$ and FeI$_2$ also both show predominantly 3R stacking (SI Fig. S14), however cross-sectional STEM images (Fig. 2k) of a small folded trilayer FeI$_2$ crystal disconnected from the main sample indicates the 1T phase, indicating high polytype plasticity, and that the presence of strain or other energetic inputs can cause spontaneous phase sliding between 3R and 1T. Monolayer regions of all 3 materials (Fig. 2a,d,g) show excellent agreement with the expected octahedral atomic structure in FeI$_2$, CoI$_2$ and NiI$_2$.

We utilise first-principles calculations to examine the origin of the structural phase transition in thin FeI$_2$, and the absence of such a transition in NiI$_2$. We first consider the magnetic configuration, calculating the energy for the non-magnetic (NM), ferromagnetic (FM), antiferromagnetic (AFM) and stripe antiferromagnetic (sAFM) structures (Fig. 2l). From the results, we confirmed that the stripe antiferromagnetic configuration[7], with pairs of alternating spin-up and spin-down magnetic moments, is the energetically favourable magnetic structure for both materials. All subsequent calculations are performed with a uniform out-of-plane spin alignment in each layer, as the small exchange energy scale (< meV) should not influence structural properties at room temperature, avoiding the need for large supercells. Comparing the calculated energy of the 1T and 3R phases for FeI$_2$ we find that the 1T phase is energetically favourable for all thicknesses, consistent with the bulk structure we identify, but not with the stacking configuration observed experimentally for few layer FeI$_2$. Nonetheless, there is only a small energy difference of just ~4.4 meV per unit cell between 1T and 3R phases at the bilayer limit, (Fig. 2j). This is an order of magnitude lower than the stacking fault penalty for NbSe$_2$ bilayers, where such faults between two competing structures are commonly observed in CVD growth[36]. This value increases to ~14 meV per unit cell for 3-4 layer FeI$_2$ (Fig.2m). The transition from 1T to the 3R stacking arrangement is therefore comparatively easy in thin samples, which enables external factors, such as strain from mechanical exfoliation or pressure from the graphene encapsulation, to induce a phase transition from the 1T phase to the 3R phase in few-layer FeI$_2$ crystals. Our calculations also show the 3R stacking arrangement of NiI$_2$ is energetically favourable for all thicknesses, consistent with both the expected structure and experimental observations for this material (Fig. 2m). However, the comparatively small stacking energy differences found in these calculations suggest that similar stacking instability may occur in few-layer NiI$_2$ as we observe in CoI$_2$ (Fig 2j-i), although we did not observe this experimentally for our samples.

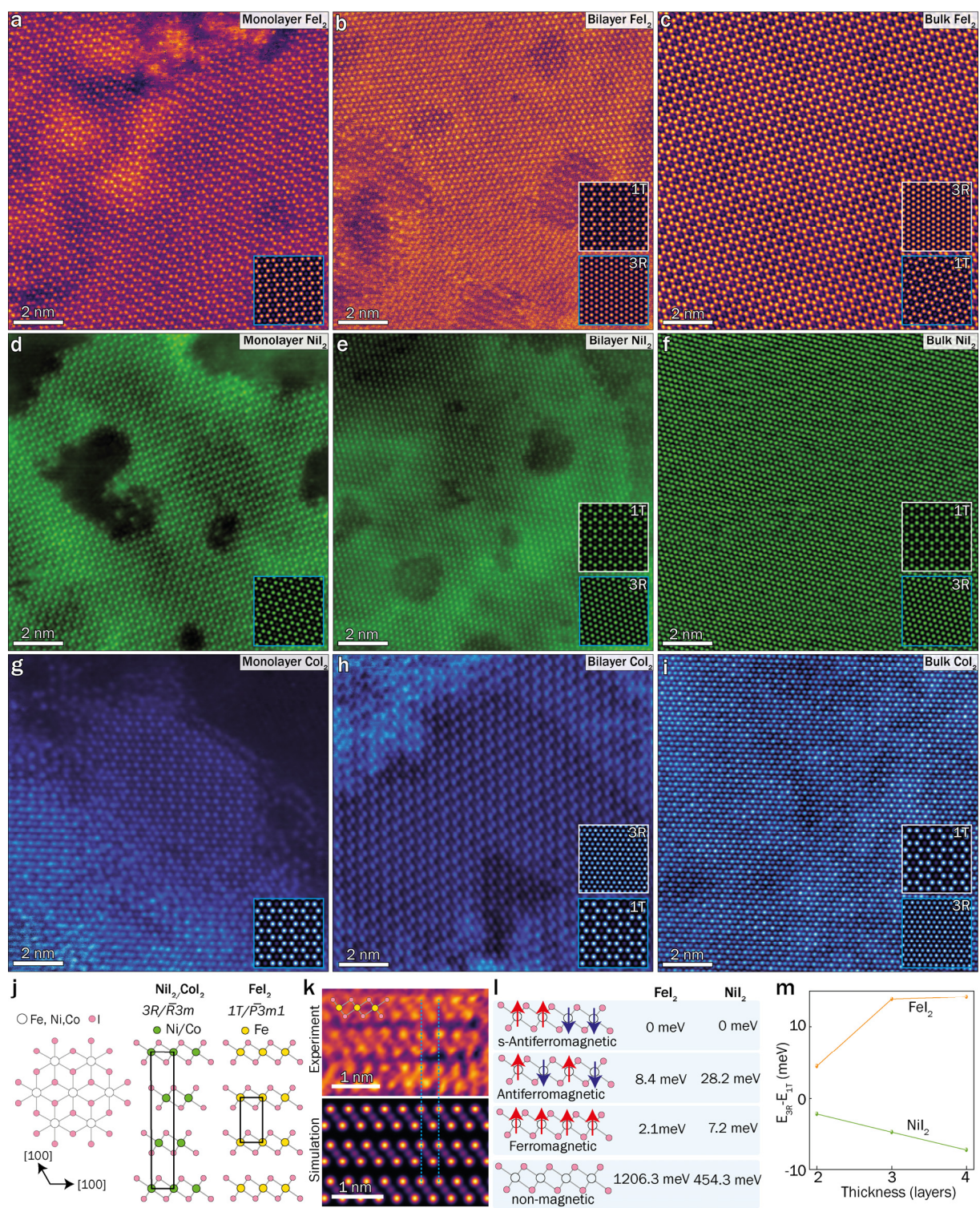

**Fig. 2 *HAADF STEM observations of layer-dependent phase changes in atomically thin MI₂ crystals.*** ***a-i,*** HAADF STEM images of monolayer, bilayer, and bulk FeI$_2$ (a-c), NiI$_2$ (d-f), and CoI$_2$ (g-i) crystal regions, respectively, with inset simulations. The inset square with the blue outline indicates which theoretical structure matches the experimental result. ***j,*** the expected structures' (top and side view) from bulk studies of FeI$_2$, CoI$_2$ and NiI$_2$, where FeI$_2$ has trigonal symmetry and space group $P\bar{3}m1$, while NiI$_2$ and CoI$_2$ have rhombohedral symmetry and space group $R\bar{3}m$. ***k.*** Cross sectional image of trilayer FeI$_2$. ***l.*** Modelling of the energy for non-magnetic (NM), ferromagnetic (FM), antiferromagnetic (AFM) and stripe antiferromagnetic (sAFM) configurations of monolayer MI$_2$, with M (Fe or Ni) atoms with no

*spin shown in grey, and different spins highlighted by red and blue arrows. **m**. Layer-dependent energetic stability of the antiferromagnetic (AFM) forms of the 3R, with reference to the stability of the 1T phase for FeI$_2$ and NiI$_2$.*

### Stability of atomic vacancies in monolayer MI$_2$

STEM is one of the few techniques able to probe the population of different types of point defect in vdW materials, which is important as the nature and quantity of point defects often determines the electronic and magnetic behaviour[37,38]. We therefore apply our atomic resolution STEM imaging to understanding the nature of atomic point defects in MI$_2$ materials, focusing on the monolayers, as these provide the clearest visualisation of the nature of defect sites (Fig. 3a-b). The population of isolated vacancy defects was quantified as a function of electron dose from comparison of HAADF STEM images to simulations (Fig. 3c). Edges and pores are excluded from this analysis and considered in a subsequent section. Each MI$_2$ layer has two iodide sublayers (above and below the metal ions) which are separately visible in plan view due to their octahedral symmetry (Fig. 2j), allowing the I$^-$ vacancy populations in the two sublattices to be separately tracked. In both MI$_2$ materials, the most common atomic point defect is I$^-$ vacancies, with one I$^-$ sublattice exhibiting a ~4 times higher density of vacancies. In both dihalides, comparatively fewer metal cation vacancies are observed (Fig. 3d). The location of point defects was tracked from frame to frame over the video series, revealing that individual vacancies often move their location in sequential images, while the total point defect density gradually decreases during extended imaging, particularly in NiI$_2$ (Fig.3d).

This suggests that defects are either highly mobile within the lattice[39] or can both form and heal over the ~2 second frame acquisition times. The decrease in areal defect density with increasing electron dose also suggests that vacancy annealing is favourable. Vacancy defects in 2D crystals can disappear either by diffusing through the lattice to edge sites or by being filled with a suitable mobile atom/ion diffusing on the surface. In graphene, hBN and MoS$_2$, aggregation of point defects during extended S/TEM imaging leads to extended pores or increasing numbers of line defects [37,40]. In our study we see no tendency for vacancy clustering to form extended defects, even after prolonged imaging (Fig. 3a-b). In fact, the total monolayer area is found to increase during imaging (SI Fig. S16), revealing that defect healing is often faster than defect creation.

To better understand the point defect population and dynamic behaviours we have performed ab-initio calculations of defect energies for both single vacancy and di-vacancy defects in FeI$_2$ and NiI$_2$ (see SI Fig. S20 and S21). We find that the formation energy for different point vacancies depends sensitively on the iodide chemical potential with iodide vacancies generally more favourable than metal vacancies in NiI$_2$ and metal vacancies more favourable in FeI$_2$. This, however, does not take into account exterior positioning of the iodide atoms in each MI$_2$ structural layer, making them more available for chemical interactions with extraneous chemical species, and hence leading to higher defect densities observed in the experiment. The agglomeration of vacancies in divacancy complexes is generally found to be an endothermic process for all species except the FeI$_2$ nearest neighbour iodide divacancy (see SI Fig S20 and S21). In all cases, the intrinsic hydrostatic strain fields around monovacancies, which drive the formation of vacancy complexes[41] from single point defects, are small in magnitude and short-ranged (<4 Å, see SI Fig S22 and S23), consistent with the low dielectric constants expected[42] in 2D materials, inhibiting their long-range interaction and subsequent agglomeration. Here, the materials are graphene encapsulated, effectively trapping degradation products in close proximity to the crystal, and providing a reservoir for vacancy healing.

The calculated vacancy formation energies can be used to estimate the threshold energy for knock on beam damage by the electron beam, based on the formulation of ref[43]. This suggests that at the experimental accelerating voltage of 80 kV, it is possible to produce iodide and metal vacancies by knock on damage in both materials. Theoretical predictions for transition metal dichalcogenides (TMDs) have shown that vacancy formation on the top sublattice is less favourable, as the displaced atom is more likely to be ''stopped'' by the other layers with top layer vacancies formed subsequently by vacancy hopping from the bottom to the top layer[44]. However, the energy difference for vacancies in the two layers in $MoS_2$ is only predicted to be ~15%, inconsistent with the 3-4 fold increase we see for iodide vacancies in one sublayer. Furthermore, electron beam damage cannot explain the reduced number of vacancies we observe over time. We therefore propose that these defects we observe are qualitatively representative of the crystal's populations of point defects after mechanical exfoliation and nanofabrication. The much larger number of $I^-$ vacancies in one sublayer compared to the other is therefore assigned to sample preparation, during which one layer is quickly covered by the protective graphene encapsulation while the other side is left unprotected for a longer period before encapsulation.

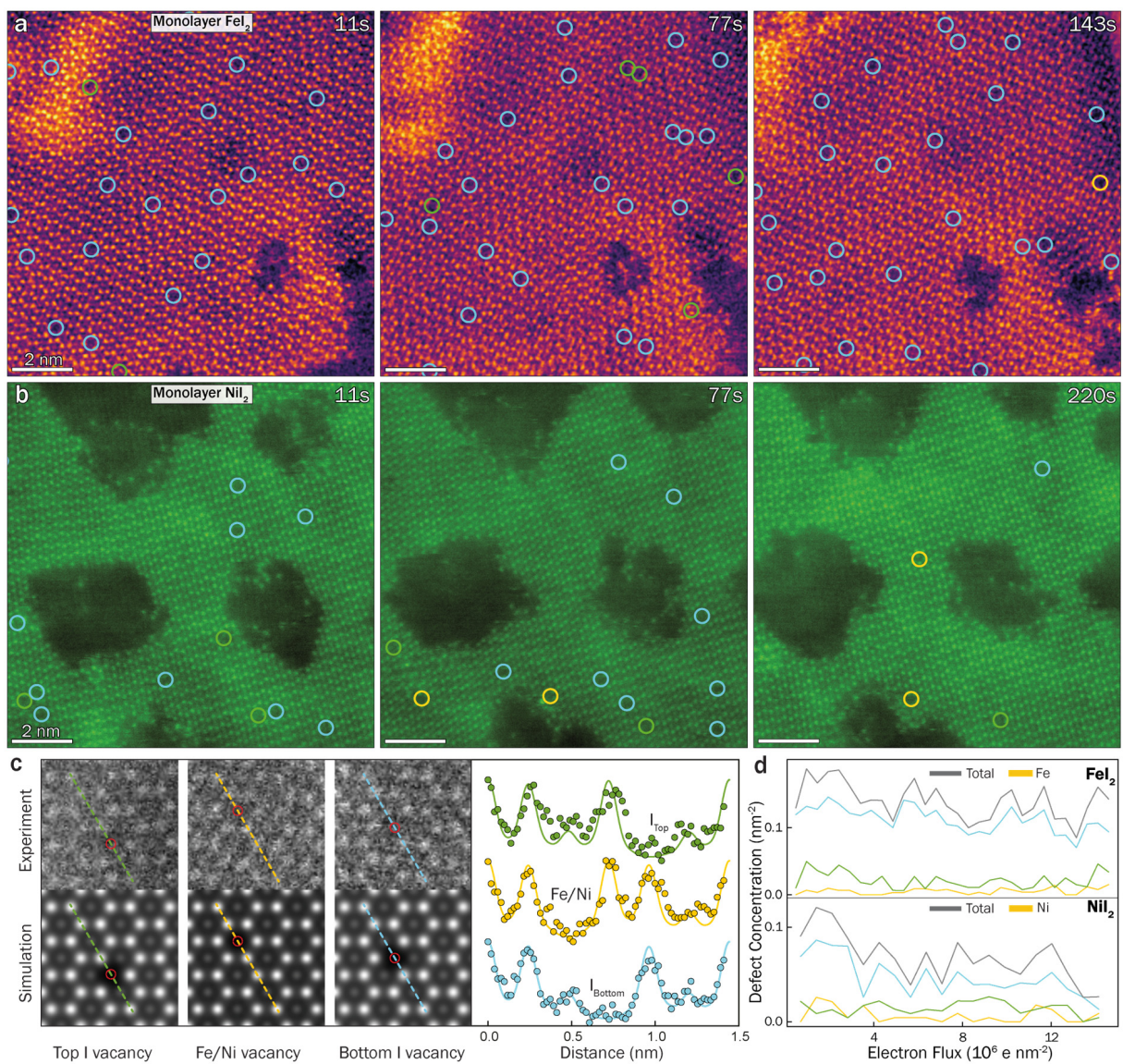

**Fig. 3 Tracking the population and migration of atomic vacancies during electron beam irradiation. a-b,** Example frames from HAADF STEM image series of FeI$_2$ and NiI$_2$ respectively. Imaging was performed with an electron fluence of 4420 electrons Å$^{-2}$s$^{-1}$. The locations of vacancy defects are highlighted by circles indicating Fe or Ni vacancies (yellow), lower sublattice I vacancies (green), and upper sublattice I vacancies (blue). **c,** Comparison of experimental images of for the three vacancy defects of FeI$_2$ to image simulations. Intensity line profiles for both experimental and simulated images are shown on the right with Fe or Ni vacancies (yellow), lower sublattice I vacancies (green), and upper sublattice I vacancies (blue). **d,** Measured areal density of vacancy defects as a function of electron flux for FeI$_2$ (top) and NiI$_2$ (bottom). Additional frames from the video series are presented in SI Fig. S15.

*Stability of edges in MI$_2$*

Edges play a critical role in charge and spin transport for 2D systems[45–47] yet engineering specific edge terminations by etching[48] or modified synthetic routes[46,49] is highly challenging. Spin-polarized edge states have recently been observed in monolayer NiI$_2$ islands grown on gold[7], the harnessing of which could provide routes to new spintronics applications. To probe the stability of different edge structures within our few layer iodides we have classified edges as either facetted (straight with a clear crystallographic termination) or curved (with no particular terminating structure). The results show that FeI$_2$ has generally curved edges with less than 20% considered facetted (Fig.4a,c) while NiI$_2$ is 30% facetted increasing to 40% after extended imaging (Fig. 4c,d). Facetted edges are dominated by a zig-zag type $<01\bar{1}0>$ layer termination, corresponding to the stoichiometric case shown in Fig. 4e. This termination differs from previous scanning tunnelling microscopy observations[7] where both Ni-rich and I-rich terminations have been observed, with the difference likely due to interaction between the NiI$_2$ and the gold substrate[50].

For high electron dose, the favourability of zig-zag edge facets for NiI$_2$, results in the increasing presence of triangular pores with 60° inner angles (Fig 4b). This suggests that the facetted structure in NiI$_2$ is significantly more favourable energetically since it involves an increased perimeter. The overall tendency of the crystal is to heal pore defects under electron beam irradiation. DFT calculations of the energy cost for the three high symmetry edge terminations illustrated in Fig.4e within the experimentally relevant range of the iodine chemical potential, $\mu_I$, are shown in Fig.4f,g. These reveal that while the stoichiometric termination is lowest energy in NiI$_2$ (consistent with experimental observations), in FeI$_2$ the stoichiometric termination competes with I-rich termination resulting in more curved edges at the atomic scale. We further consider the competition between edges and the adatom defects of a metal atom atop an iodide site (M@I$_{top}$) and iodide on a metal side (I@M), which are the lowest-energy configurations among the three possible symmetric adsorption sites and provide a conservative reference energy for edge growth. Fig.4f,g show that for FeI$_2$ these are always less energetically favourable than the stoichiometric edge, while at low $\mu_I$ it can be more energetically favourable to have a Ni@I$_{top}$ adatom. This reinforces the high stability of the stoichiometric edge in NiI$_2$ when it is competing with adatom defects.

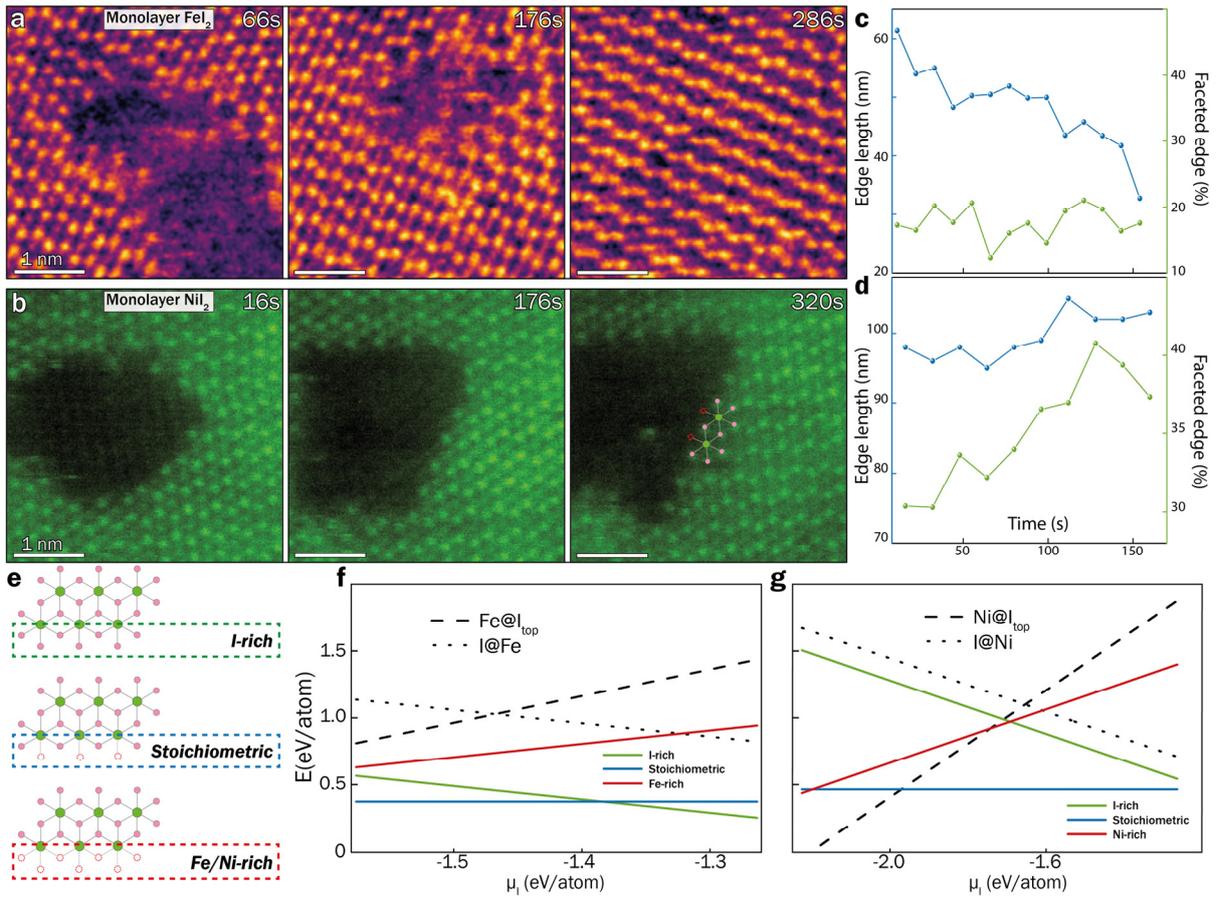

***Fig. 4 Dynamics of extended edge defects in MI$_2$ and edge formation energy. a,b,** Evolution of FeI$_2$ (a), and NiI$_2$ (b) edge defects during continued electron irradiation under a constant fluence of 4420 electrons Å$^{-2}$s$^{-1}$. **c,d,** Edge length and faceted edge percent evolution under electron irradiation for FeI$_2$ (c) and NiI$_2$ (d). **e,** The schematics of three types of edges in MI$_2$. **f,g,** Formation energy of different edge structures as a function of the iodine chemical potential for FeI$_2$ (f) and NiI$_2$ (g). The iodine reference chemical potential, $\mu_I$ is taken in the range $E_{coh,I} + \Delta H_f^{MI_2}/2 < \mu_I < E_{coh,I}$, where lower and upper bounds correspond to equilibrium of M, I in MI$_2$ with bulk metal and bulk iodine, respectively. The formation energies of different adatom defects are also shown, where M@I$_{top}$ is a metal atom atop an iodide site and I@M is an iodide on a metal side.*

*Conclusion*

The structural characterisation of transition metal diiodides performed here has revealed a sensitivity to sample fabrication conditions uncommon among more widely studied 2D materials. The small energy penalty for alternative stacking configurations in the few layers case suggests modest strain or pressure during encapsulation can result in alternative polytypes to those observed in the bulk crystal, with significant implications for the magnetic properties at the thin limit, and the potential to exploit this flexibility to tune interlayer interactions and exploit sliding transitions. Defect dynamics are also significantly different to those reported for TMDs, hBN and graphene with no evidence of defect clustering for prolonged STEM imaging. DFT simulations suggest this may be due to weak defect induced strain fields and unfavourable vacancy aggregation energies, opening the possibility of using defect engineering to create disordered spin/magnetic textures in thin di-iodides without the risk of pore formation.

More broadly, our results demonstrate that characterisation of atomic scale phenomena in even the most air-sensitive 2D materials down to the monolayer limit is both possible and practical, if they can be hermetically encapsulated with no polymeric or air deposited residue inside the inert envelope. Our polymer free fabrication strategy based on silicon nitride cantilevers can simply and rapidly create such structures in both high vacuum and inert glovebox environments with high-yield, opening the possibility of extending advanced atomic resolution characterisation of monolayer materials to even the most environmentally sensitive 2D crystals.

**Methods**

**Crystal growth**:

$FeI_2$ crystals were made by direct reaction of the two elements in a quartz ampoule using the chemical vapor transport (CVT) method. In the ampoule (50 x 250 mm) were placed iron (99.9%, -100 mesh, Strem, USA) and iodine (>99.9%, Fisher Scientific, USA) corresponding to 25 g of $FeI_2$. Iodine was used in 1 at.% excess towards stoichiometry. The ampoule was melt sealed under high vacuum (< $1 \times 10^{-3}$ Pa using a diffusion oil pump and liquid nitrogen trap) while the charge was also cooled under liquid nitrogen. The ampoule was then placed in a crucible furnace and heated gradually to 500 °C while the top part was cooled to 200 °C. After four days of reaction the $FeI_2$ was formed. The ampoule was then placed in two zone furnace for CVT growth. First the growth zone was heated to 600 °C while the source zone was kept at 400 °C. After 50 hours the thermal gradient was reversed, and source zone was heated to 560 °C while the growth zone was kept at 480 °C for 14 days. The ampoule was then cooled to room temperature and opened in argon filled glovebox.

To prepare $NiI_2$ by CVT the ampoule (50 x 250 mm) contained nickel (99.99%, -100 mesh, Abcr Germany) and iodine (>99.9%, Fisher Scientific, USA) corresponding to 25g of $NiI_2$. Iodine was again used in 1 at.% excess towards stoichiometry and the ampoule was melt sealed under high vacuum similar to $FeI_2$. The ampoule was then placed in a crucible furnace and heated gradually to 500 °C while the top part was cooled below 200 °C. After four days $NiI_2$ was formed. The ampoule was then placed in two zone furnace for CVT growth. First the growth zone was heated to 750 °C while the source zone was kept at 500 °C. After 50 hours the thermal gradient was reversed, and source zone was heated to 750 °C while the growth zone was kept at 680 °C for 14 days. The ampoule was then cooled to room temperature and opened in argon filled glovebox.

**STEM sample fabrication**:

Graphene flakes were exfoliated onto Si/SiO$_2$ substrates using standard mechanical exfoliation techniques and characterized under an optical microscope to identify those with suitable size and thickness. Fresh holey SiN$_x$ cantilevers were prepared with a tri-layer metal coating: 1 nm Cr as the adhesion layer, 5 nm Pt to ensure surface flatness, and 1.5 nm Au to enhance adhesion between graphene and SiN$_x$. Graphene flakes were picked up by the holey cantilevers at 150 °C in air within 30 minutes of the metal coating to prevent contamination or oxidation of the fresh Au. The graphene/SiN$_x$ cantilever assembly was then annealed under vacuum at 200 °C for 1 hour to stabilize the graphene on the cantilever.

At the same time, MI$_2$ flakes were exfoliated onto Si/SiO$_2$ substrates inside an argon glovebox with O$_2$ and H$_2$O levels maintained below 0.1 ppm. These flakes were characterized using an optical microscope to identify thin, uniform regions suitable for subsequent assembly. The selected MI$_2$ flake was picked up using the graphene/SiN$_x$ cantilever, then a second graphene layer was picked up to achieve full encapsulation of the MI$_2$. The bottom graphene was chosen to be smaller than the top layer but larger than the MI$_2$ flake to ensure easy pick up and complete encapsulation of the heterostructure. Finally, the assembled holey SiN$_x$ cantilever with the graphene/MI$_2$/graphene heterostructure was mounted onto a custom-designed TEM grid in air.

**STEM Imaging and Analysis**:

High resolution STEM imaging was performed with a JEOL ARM300CF double aberration-corrected microscope with a cold FEG electron source operated at an accelerating voltage of 80 kV and probe current of 8 pA, using a probe convergence angle of ~21mrad and detector collection angle of 73-155 for HAADF STEM image acquisition. STEM simulations were performed using open-source python package abTEM[51] matching the experimental parameters of the JEOL ARM300CF. Image analysis and filtering was performed using open-source python package Hyperspy[52] version 2.3.0.

**Density Functional Theory calculations**:

Spin-polarized density functional theory (DFT) calculations were performed using the VASP package, with computational parameters as outlined in Supplementary Section 9. While magnetic configurations are markedly favoured over non-magnetic ones, the relatively weak intralayer exchange values, consistent with the low Néel temperatures in FeI$_2$ (T$_N$ = 11 K[53]) and NiI$_2$ (T$_N$ = 76 K[54]), suggests spin disorder and an overall paramagnetic behaviour at high temperatures. Consequently, calculations of point defects and edges were carried out using fully spin-polarised structures. Monovacancies and divacancies were created by removing the corresponding number of atomic species from a periodically repeated 4×4 or 6×6 supercell, while different edge conformations were created in small ribbon-geometries (periodically repeated in one direction), with the required edge termination. All structures were then fully structurally relaxed and defect formation energies were calculated as a function of the iodine chemical potential, $\mu_I$, for values in the range

$$E_{coh,I} + \frac{\Delta H_f^{MI_2}}{2} < \mu_I < E_{coh,I},$$

where $E_{coh,I}$ is the cohesive energy of bulk iodine and $\Delta H_f^{MI_2}$ is the enthalpy of formation of bulk MiI$_2$ compounds.

**Data Availability:** Raw images and underlying data are available to download [link to be provided upon manuscript acceptance]. No custom code has been used in this work.

**Acknowledgements**: The authors acknowledge funding from the European Research Council (ERC) under the European Union's Horizon 2020 research and innovation programme (Grant ERC-2016-STG-EvoluTEM-715502 and ERC-2020-COG-QTWIST-101001515). We also thank the Engineering and Physical Sciences Research Council (EPSRC) for funding under grants EP/Y024303, EP/S021531/1, EP/M010619/1, EP/V026496/1, EP/V007033/1, EP/Z531121/1, EP/S030719/1, EP/V001914/1, EP/V036343/1 and EP/P009050/1, and also for the EPSRC Centre for Doctoral Training (CDT) Graphene-NOWNANO. TEM access was supported by the Henry Royce Institute for Advanced Materials, funded through EPSRC grants EP/R00661X/1, EP/S019367/1, EP/P025021/1 and EP/P025498/1. RG. and VF. acknowledge funding from the European Quantum Flagship Project 2DSIPC (no. 820378). We thank Diamond Light Source for access and support in use of the electron Physical Science Imaging Centre (Instrument E02 and proposal numbers MG39088) that contributed to the results presented here.

**Author contributions:** R.G. and S.J.H. conceived the study. W.W., N.C., S.S.A. developed and tested the holey $SiN_x$ membrane transfer. G.T. and S.J.H. performed electron microscopy characterisation with help from W.W.,S.S.A. J.M., X. L. and V.F. performed the theory calculations. F.S. performed AFM measurements. R. Z. performed the SEM measurement. S.Z., O.C. supplied high quality $MI_2$ crystals. R.G., S.J.H., W.W., and G.T. wrote the manuscript. All authors contributed to the discussions and commented on the manuscript. W.W., G.T. have contributed to this work equally.

**Competing interests:** The authors declare no competing interests.